\documentclass[twocolumn]{aastex631}
\usepackage{amsmath}
\usepackage{graphicx}	
\usepackage{chemformula}
\usepackage{verbatim}
\usepackage{hyperref,natbib,amsmath,color,ulem}
\usepackage{soul}

\usepackage{color,epstopdf,placeins}
\definecolor{darkgreen}{rgb}{0.2, 0.7, .2}

\def\g{\, \rm g}
\def\K{\, \rm K}

\def\s{\, \rm s}
\def\km{\, \rm km}
\def\cm{\, \rm cm}
\def\m{\, \rm m}
\def\dyne{\, \rm dyne}

\definecolor{darkgreen}{rgb}{0.2, 0.7, .2}
\newcommand{\x}{\sout}
\newcommand{\z}{\color{blue}}
\newcommand{\w}{\color{black}}
\newcommand{\y}{\color{red}}
\newcommand{\atc}{\color{red}}

\shorttitle{Eccentric  Disk}
\shortauthors{Goksu, Kutra and Wu}

\graphicspath{{./}{figures/}}

\usepackage{orcidlink}

\begin{document}

\title{On the Rigidly Precessing, Eccentric Gas disk Orbiting the white dwarf SDSS J1228+1040}
\author[0009-0002-4988-2545]{Ates  Goksu}
\affiliation{David A. Dunlap Department of Astronomy \& Astrophysics, University of Toronto}
\author[0000-0002-7219-0064]{Taylor Kutra}
\affiliation{David A. Dunlap Department of Astronomy \& Astrophysics, University of Toronto}
\affiliation{Lowell Observatory}
\author[0000-0003-0511-0893]{Yanqin Wu}
\affiliation{David A. Dunlap Department of Astronomy \& Astrophysics, University of Toronto}

\begin{abstract}
Metal pollution onto white dwarfs is a wide-spread phenomenon that remains puzzling. Some of these white dwarfs also harbour gaseous debris disks. Emission lines from these disks 
open a unique window to the  physical properties of the polluting material, lending insights to their origin. We model the emission line kinematics for the gas disk around SDSS J1228+1040, a system that has been monitored for over two decades. We show that the disk mass is strongly peaked at $1 R_\odot$ (modulo the unknown inclination), and the disk eccentricity decreases from a value of $0.44$ at the inner edge, to nearly zero at the outer edge. This eccentricity profile is exactly what one expects if the disk is in a global eccentric mode, precessing rigidly
under general relativity and gas pressure. The precession period is {\w about two decades.} {\w We infer that the} mass of the gas  disk is roughly equivalent to that of a {\w 50}-km rocky body, while the mass of the accompanying dust disk is likely insignificant. The disk eccentricity confirms an origin in tidal disruption, {\w while the short disk diffusion time suggests that the disruption event happened a few centuries ago. Moreover, we argue that the initial orbit for the disrupted body, and that of its putative planetary perturber, fall within an AU around the white dwarf.}
{\w The total mass of the source population is likely orders of magnitude more massive than our own Asteroid belt, and does not seem to exist around main-sequence stars.}
\end{abstract}


\section{Introduction} \label{sec:intro}


About a third of all white dwarfs show signs of on-going or recent accretion of heavy metals  \citep[e.g.,][]{Zuckerman2003}.  And starting from the first example of 
white dwarf G29-38 \citep[][]{1987ApJ...322..852K,1990ApJ...357..216G, Jura2003}, many are now known to also exhibit infrared excesses, signs of circumstellar dust disks \citep[e.g.][]{2005ApJ...632L.115K, 2006ApJ...646..474K, 2009ApJ...694..805F, 2011ApJS..197...38D, 2012ApJ...759...37D}. 
These disks are likely  metallic in composition and are responsible for the pollution. It is now commonly believed that large asteroids (and/or comets) around these stars are, for some reason, excited to high eccentricities and are tidally disrupted when they approach the white dwarfs within the Roche radius. The resulting debris forms the dust disk. However, many key elements in this story, including the source and the orbital excitation for these bodies, remain mysterious  {\w \citep[for reviews, see][]{2016NewAR..71....9F, Veras2016,Veras2021}. }

Interestingly, a few percent of these dusty white dwarfs are also known to possess gaseous debris disks \citep{Manser2020}.
First discovered from SDSS data by \citet{2006Sci...314.1908G} around the white dwarf 
SDSS J122859.93+104032.9 (short-named as J1228 below), about two dozens such disks are now known 
\citep{2006Sci...314.1908G, 2007MNRAS.380L..35G, Gansicke2008,2012MNRAS.421.1635F,Fusillo2021}. These disks manifest as double-peaked emission lines in the spectra, most conspicuously in Ca II infrared triplets. They are found exclusively around white dwarfs hotter than $\sim 13,000\K$, likely because only such stars are {\w luminous} enough to sublimate dust at a distance of $\sim 1 R_\odot$. {\w The rarity of gaseous disk is likely explained by the rarity of these hot stars (constituting only a few percent of all white dwarfs).}

Unlike dust disks which reveal little about their kinematics, compositions, or density distributions \cite[e.g., see a discussion in][]{Nixon2020}, gaseous disks open a lucky window. 
The characteristic double-peaked emission lines from these disks contain information about  their radial extent, eccentricity,  surface density and temperature profiles. Interestingly, {\w some of them } exhibit asymmetric lines that are most easily interpreted as the signature of an eccentric disk. 
Moreover, these disks also appear to be time variable {\w \citep[e.g.,][]{2014MNRAS.445.1878W, 2015ASPC..493..279W,Dennihy2018,Manser2021}.} 
For instance, J1228, the best monitored system \citep{Manser2016,2019Sci...364...66M}, shows gradual variations of its line profile over a timescale of decades. This is much longer than the local Keplerian timescale, which is of order hours. 

This kinematic information offers the hope of understanding the origin of these debris disks. Here, we undertake a study of the J1228 gaseous disk, with an aim to answer the following specific questions. 

First, how does the disk manage to retain its eccentric shape? This disk is known to extend radially {\w (with a width $\Delta r \sim r$)}. 
If so, general relativistic effects precess the gas differentially,  and would have led to its total circularization within a few decades. 
In fact, such a consideration motivates \citet{Hartmann2011,Metzger2012,Cauley2018,Fortin2020} to propose that the observed line profiles are not due to an eccentric disk, but are instead due to non-axisymmetric brightness patterns (a vortex, or a spiral wave) on a circular disk. {\w An alternative possibility is laid out by \citet{2018ApJ...857..135M} wherein they argue that gas pressure in the disk can resist GR and maintain the disk in a state of rigid precession.} We resolve this issue by first fitting a physically motivated disk model (\S \ref{sec:Model}) against detailed observations (\S \ref{sec:Method}), and then 
{\w show that the disk is indeed in a state of pressure-maintained rigid precession} (\S \ref{subsec:precession}), {\w as is proposed by \citet{2018ApJ...857..135M}.}

Second, {\w the kinematic information we extract from the emission lines allows us to set unique constraints on the origin of the gas disk, and more generally on the origin of white dwarf pollution} (\S \ref{sec:insights}).

\section{Physical Model} \label{sec:Model}

We will construct a Keplerian disk model to reproduce the observed double-peaked line profiles in the Ca II infrared triplet.  {\w We assume that the gaseous disk region is free of solid grains. As we argue below (\S \ref{subsec:insights4}), one should not expect dust to co-exist with an eccentric gas disk.}

In such an exercise, as one only measures the line-of-sight velocity, and is ignorant  of the orbital period,  one can only determine the length combination, $a/\sin^2 i$, where $i$ is the orbital inclination {\w relative to the line of sight (with $i=0$ being face-on)}. This differs from the usual radial velocity literature where one also knows the orbital period and can determine $a\sin i$. We suppress the factor $1/\sin^2 i$  in this section, but re-introduce it in later discussions. 

\begin{figure*}[!ht]
    \centering
    \includegraphics[width=0.9\textwidth]{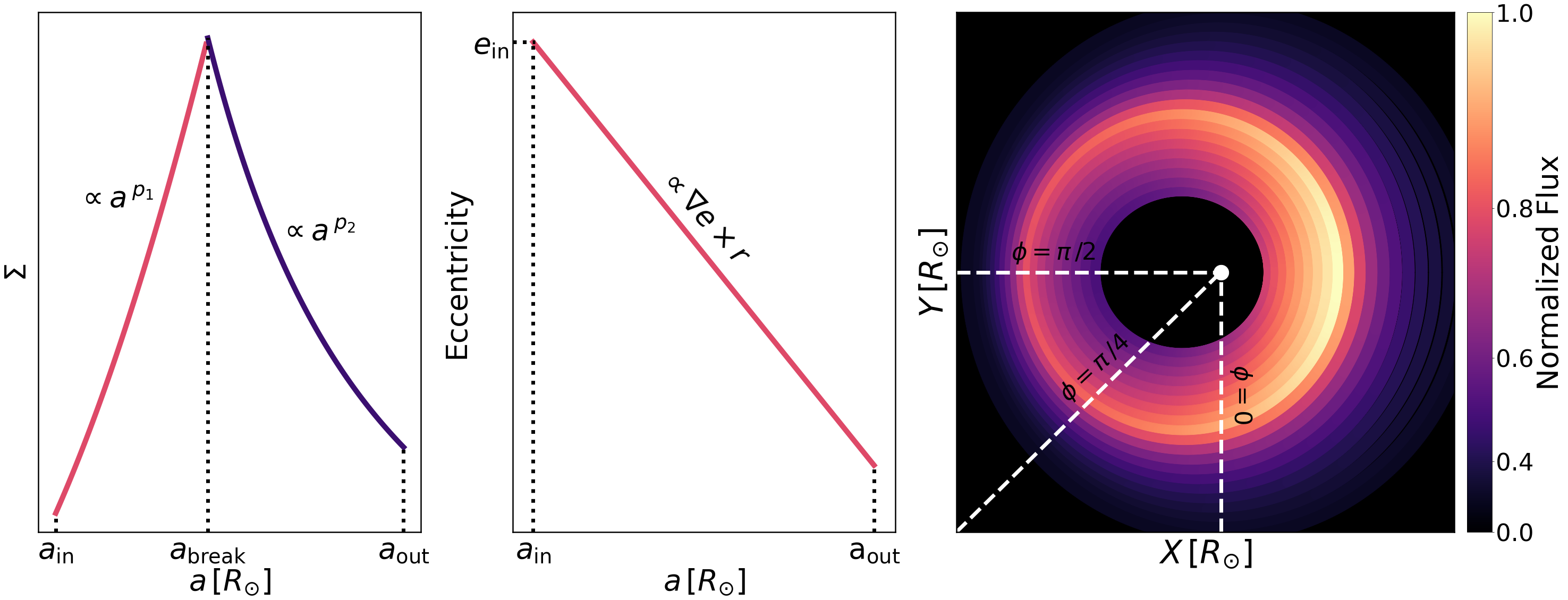}
    \caption{An illustration of our disk model. We represent the disk using a discreet set of elliptical, aligned, Keplerian rings, with a surface density profile (left panel) that is composed of two broken power-laws,  and an eccentricity profile (middle panel) that is linear in radius. The right panel is a birds-eye view of the disk with the colour indicating normalized emission. This figure is produced using our best fit parameters. The phase angle $\phi$ takes the value of $0$ if the disk long-axis lies on the plane of the sky.
    }
     \label{fig:modeldisk}
\end{figure*} 

Furthermore, in order to translate line flux into gas density,  we need to consider the physics of emission. 
The Ca II triplet  are most likely recombination lines, namely, spontaneous emission from Ca II ions at excited levels, arriving after photo-ionization and recombination. The line fluxes should, therefore, scale linearly with the rate of  recombination, which, at equilibrium, equals the rate of photo-ionization. So the emissivity should scale as $\propto n_\gamma\, n_{\rm Ca II}$, where $n_\gamma$ is the number density of ionizing photons from the white dwarf, and $n_{\rm Ca II}$ that of Ca II. The above scaling remains valid even when the disk is very optically thick to the recombination lines (as is the case for our disk). 

This consideration allows us to determine the local gas density, under some simplifying assumptions. In particular, we will assume that $n_\gamma$ depends on the radial distance from the white dwarf as $n_\gamma \propto r^{-q}$, with $q$ being a free parameter. We expect $q= 2$ when the disk is optically thin to the ionizing photons, {\w while $q > 2$ when the radial optical depth to these photons is larger than unity}. We will also assume that the disk has a negligible vertical extent {\w ($H$)} and is not being viewed nearly edge-on {\w (i.e., $H \ll r \cos i$)} These allow us to determine the local column density from the  {\w local} height-integrated emission, {\w without worrying about the disk vertical structure.}
Lastly, we assume that Ca II perfectly traces the local gas density.  

We model the gas disk around J1228 as an assembly of $20$ tightly packed, co-focal elliptical Keplerian rings. Their semi-major axes are evenly distributed between $a_{\rm in}$ and $a_{\rm out}$. The  gas surface density profile is assumed to be a broken power-law with a transition radius $a_{\rm break}$:
\begin{equation}
    {\Sigma} \propto 
    \begin{cases}
        a^{p_1} \, , \,\hskip0.5in a_{\rm in} \le a < a_{\rm break}\, ; \\ 
        a^{p_2}\, , \hskip0.5in a_{\rm break} \le a \le a_{\rm out}\, .
    \end{cases}
    \label{eqn:densrad}
\end{equation}
Such a broken profile is motivated by the observed line profiles, which also suggest that  $p_1 > 0$ and $p_2 < 0$. 

As the simplest approximation, we assume that the ring eccentricities vary linearly with the semi-major axis as
\begin{equation}
   e(a) = e_{\rm in} + {{de}\over{da}} \times (a-a_{\rm in}) \label{eq:defecc}\, .
\end{equation}
We do not specify the sign of the eccentricity gradient ($de/da$).
The rings are assumed to remain apse-aligned at all times. This requires the disk to precess rigidly,  a working assumption we justify in \S \ref{subsec:precession}.  

The line emissivity $\epsilon$, from  a ring segment of length $d\ell$, is therefore 
\begin{equation}
  \epsilon = {\rm const} \times \, 
     r^{-q} \, \times \Sigma 
    \,\times  \left (\frac{\frac{1}{v}{d\ell}}{\oint {\frac{1}{v} d\ell}} \right ) \, ,\label{eqn:emission}
\end{equation} 
with $v$ being the Keplerian velocity of the segment. 
The first factor ($r^{-q}$) describes the radial dependence for the ionizing flux, and 
the last factor describes the fractional mass within the line segment. This scales inversely with the local Keplerian velocity as mass is conserved along a Keplerian streamline. This behaviour is behind the so-called  `apocentre glow' in debris disks\citep{2016ApJ...832...81P,2017ApJ...842....8M}. The overall normalization constant is discussed in  \S \ref{sec:insights}, where we show that 
our model disk {\w can also account for the total observed flux, in addition to the line profile}.

To assign a Doppler velocity to the above line segment, we introduce a phase angle $\phi$, where $\phi = 0$ corresponds to the case where the orbital long axis lie on the plane of the sky. The resultant line profile is symmetric at this phase, while the line is at its most asymmetric when $\phi = \pi /2$. 

All our model parameters are illustrated in  
Fig. \ref{fig:modeldisk}.
In the following, we proceed to search for the best model parameters for the  observed Ca-II line data from \citet{Manser2016}.

\section{MCMC And Results} \label{sec:Method}

\subsection{Markov Chain Monte-Carlo} \label{mcmc}\

\citet{Manser2016} have gathered spectra of J1228 from March 2003 to May 2015 in a total of 18 epochs, using the VLT, the SDSS telescope and the WHT. C. Manser has kindly provided us with the data. Some subsequent observations are presented in \citet{2019Sci...364...66M} but are not used for fitting.

To prepare the emission profiles for analysis, we convert the  Ca-II triplet data from wavelength (\AA) to velocity (${\rm km} \, {\rm s}^{-1}$), using the atomic rest-wavelengths, 
and a systemic velocity of $+22 \km/\s$. This value is within the range reported by \citet{Manser2016}: $+19 \, \pm 4  \, {\rm km} \, {\rm s}^{-1}$.\footnote{One can also determine the systemic velocity from the white dwarf spectra, after accounting for a gravitational redshift of $35\km/\s$ for lines emitted from the surface of the white dwarf.}
It is chosen so that the emission lines at the June 2007 epoch, which have very similar amplitudes in the blue- and red-shifted peaks,  are also symmetric in the velocity space.  We co-add the three lines to produce a joint line profile.

The precession of the disk means we have the good fortune to observe it from different vantage points, each giving some unique constraints on the disk model. We decided, initially, to focus on data from three (equally spaced) epochs: June 2007, June 2011, and May 2015. We assign a phase of  $\phi = \pi/2$ to the first epoch (most symmetric), and phases of $0.31$ and $-0.92$ to the remaining two. This choice is motivated by our inferred precession period of $\sim 20 \, {\rm yrs}$ (see below). Our initial attempt did not produce a satisfactory fit to the June 2007 data, so we proceed to include two more epochs, April 2007 and July 2007, with their corresponding $\phi$ values, into the procedure. This serves to strengthen the model constraint around $\phi = \pi/2$. 

We now determine the best fit model parameters, using the \textit{emcee} \citep{2013PASP..125..306F} implementation of the  Markov Chain Monte-Carlo (\textit{mcmc}) method.

This procedure requires appropriate priors. For each of our 8 parameters, we choose a flat prior over a wide range (Table \ref{tab:bestfit}).  
Our prior on the eccentricity profile warrants some comments. First, we posit that $e \in [0,1)$ everywhere. {\w Second, we insist that streamlines cannot cross within the disk. 
As streamline crossing likely occurs at highly supersonic speed, the resultant shock will remove much of the orbital energy from the gas and cause rapid in-spiral, effectively truncating the disk \citep[i.e.][]{Artymowicz1994}. This leads us to} impose the  condition
\citep{GoldreichTremaine}
\begin{equation}
   \bigg |e(a) + \frac{\mathrm{d} \, e(a)}{\mathrm{d} \, \mathrm{ln}(a)} \bigg | 
   < 1\, .
   \label{eqn:streamline}
\end{equation}

\begin{figure}[h!]
    \centering
    \includegraphics[width = 0.45\textwidth,clip,trim=0 80 0 120]{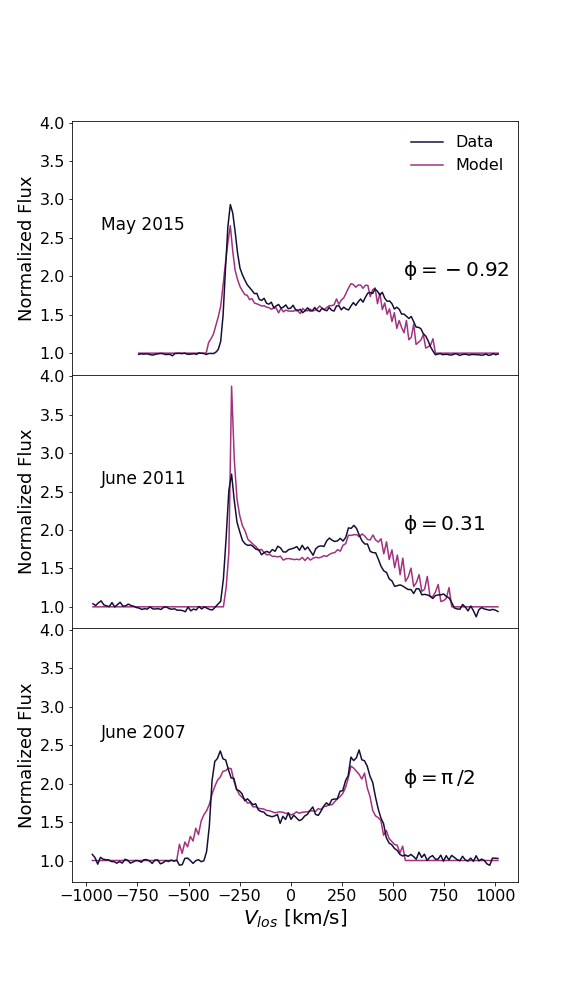}
    \caption{Comparing the emission profiles between model and data,  for three phases. At $\phi=\pi/2$, we expect symmetry but the data shows a deficit of emission in the blue wing.}
    \label{fig:epochs}
\end{figure}

\begin{figure}[h]
    \centering
    \includegraphics[width=\linewidth]{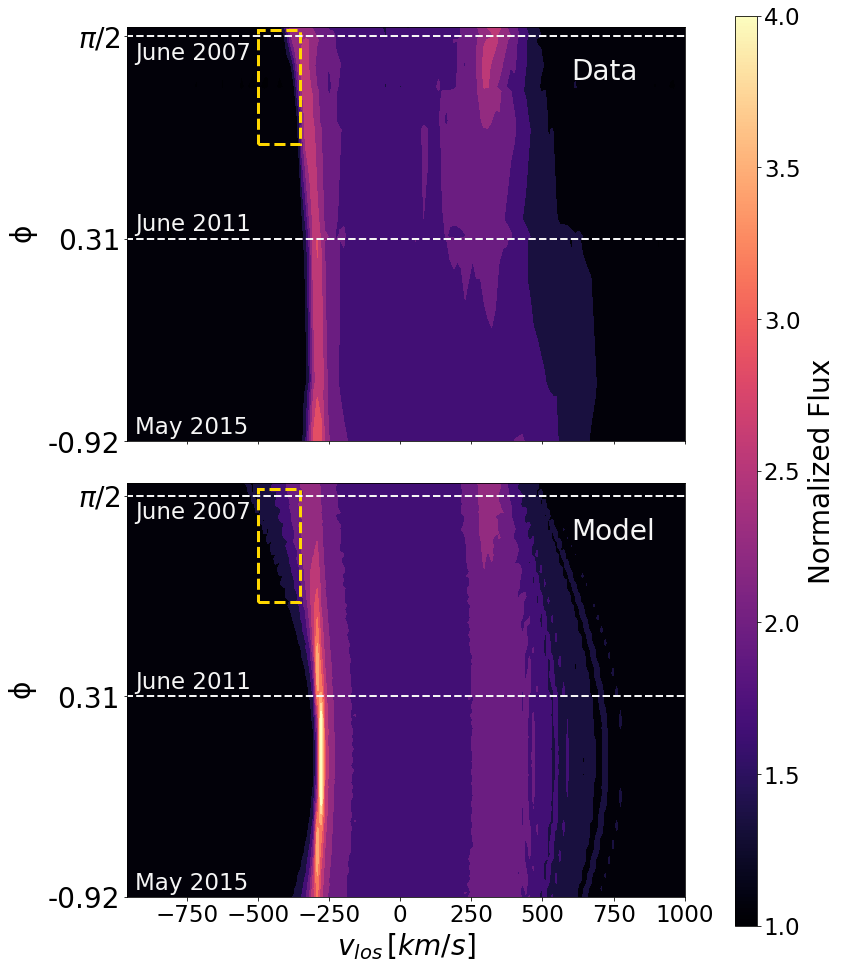}
    \caption{
    Observed and modelled emission profiles as functions of the precession phase (vertical axis). The data plot is generated using epochs from April 2007 to May 2015  \citep{Manser2016} and summing over the Ca II infrared triplet. 
    While the overall resemblance is good, we highlight one notable exception. Near phase $\phi = \pi /2$, the observed profile is less symmetric than the model and shows a sharper cutoff at the blue peak (yellow boxes).}
    \label{fig:riverplot}
\end{figure}

We run {\it emcee} with 100 walkers and iterate each for 4000 steps. This ensures that the auto-correlation time is a sufficiently small fraction of the total run. We then trim the first 2000 steps to minimize the effects of the initial conditions. The full results are presented in Figure \ref{fig:corner}. There is a good convergence for all parameters, though some show a slight (but unimportant) bimodal distributions in their posteriors.
The maximum-likelihood parameters, and their corresponding uncertainties, are presented in Table \ref{tab:bestfit}. Furthermore, the resultant line profiles for the three chosen epochs are presented in Figure \ref{fig:epochs}; while Fig. \ref{fig:riverplot} illustrates them for a  continuum  
of phases. 

While the overall comparison is satisfactory, we note that the data show a conspicuous shortage of emission in the blue wing at $\phi = \pi /2$ (bottom panel in Fig. \ref{fig:epochs}, yellow box in Fig. \ref{fig:riverplot}). This is the phase where we expect symmetric emission, and indeed the blue and the red peaks do look symmetric. The deficit is only in the wing, and it persists in observations from nearby epochs (April and July 2007). The latter rules out the possibility that the choice of our $\phi = \pi/2$ epoch is the cause of such an asymmetry. We have no explanation for this deficit.

\subsection{Disk Properties}
\label{sec:results}

\begin{figure*}
    \centering
    \includegraphics[width=0.9\linewidth]{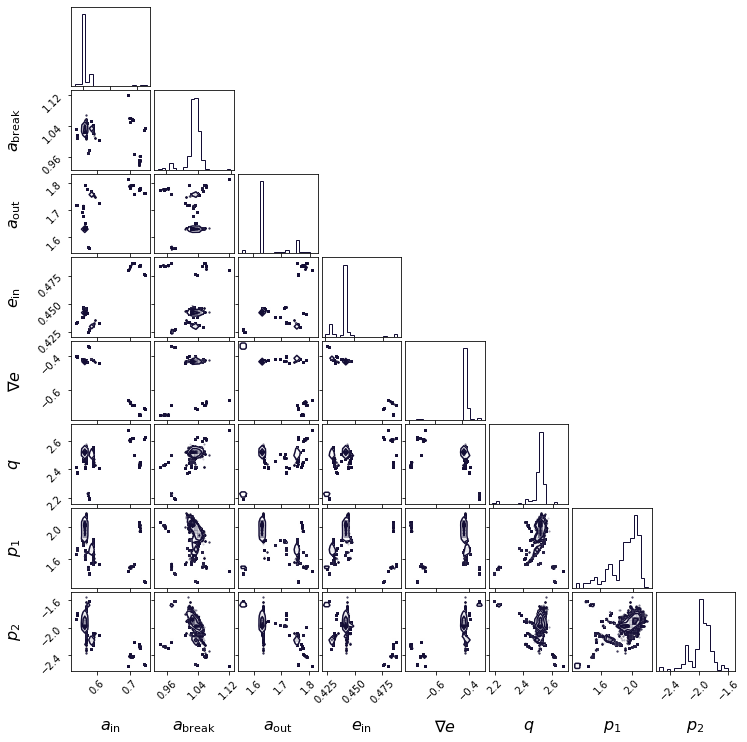}
    \caption{Corner plot for the 1- and 2-D marginalized 
 posterior distributions for our model parameters, with the first 2000 steps discarded as burn-in. We see good overall convergence for our parameters.}
    \label{fig:corner}
\end{figure*}

We now review the properties of our best-fit model. 
\citet{2006Sci...314.1908G} and \citet{Hartmann2016} have previously determined the inner and outer radii for the gas disk: $a_{\rm in} \sim 0.6$ $R_\odot$ and $a_{\rm out} \sim 1.2 \, R_\odot$. Our solutions are broadly consistent with their values, with $a_{\rm in} = 0.57 R_\odot$ and $a_{\rm out} = 1.7 R_\odot$. Bear in mind that the physical lengths are smaller than these by $\sin^2 i$. 
Our value for the inner eccentricity ($e_{\rm in}=0.44$) is also consistent with that inferred by 
\citet{Manser2016, 2019Sci...364...66M}
This value is higher than the $e=0.021$ value in the original discovery paper \citep{2006Sci...314.1908G}, because they happened to catch the lines when they were more symmetric.

Our most interesting result {\w from the MCMC fit} is the eccentricity gradient. We find  a significant and negative eccentricity gradient: $d e/da = -0.42 \pm 0.059$. Compared to the inner disk, the outer disk is substantially more circular, and is  in fact consistent with being circular. This result can be intuitively understood by looking at the top panel of Fig. \ref{fig:epochs}: the sharp spike in  the blue wing comes about because the rings are compressed together at apoapse (both in physical space and in velocity space). To do so, the inner part of the disk has to be more eccentric. {\w A negative eccentricity gradient is a characteristic feature of a rigidly precessing gas disk \citep{2018ApJ...857..135M}. }
We return to this  gradient in \S  \ref{subsec:precession}.

\begin{deluxetable}{cccc}\label{tab:bestfit}
\tablecaption{Most likely model parameters and their $1 - \sigma$ uncertainties.}
\tablehead{
\colhead{Parameter}
&  \colhead{Prior}
&\multicolumn{2}{c}{Solution} \\ 
 \colhead{}   
 & \colhead{} 
 & \colhead{Mean ($\mu$)}
 & \colhead{Uncertainty ($\sigma$)}}
\startdata
{$a_{\rm in} \, [R_\odot]$}
& $\in [0.3, 4]$ & {0.57}     & $5.5 \%$ \\ 
{$a_{\rm break}\, [R_\odot]$}    & $\in [a_{\rm in}, a_{\rm out}]$ & {1.0} & $2.8 \%$ \\
{$a_{\rm out}\, [R_\odot]$}      & $\in [a_{\rm in}, 7]$ & {1.7} & $4.1 \%$ \\ 
{$e_{\rm in}$} & $\in [0, 1)$ & {0.44} & $3.9 \%$ \\ 
{$\nabla e$} & Eqn. \ref{eqn:streamline} & {-0.42} & $14 \%$ \\ 
{$q$}  & $ > 0$ & {2.4} & $6.8 \%$ \\
{$p_1$}  & $\in (0, 5)$ & {1.8} & $17 \%$ \\ 
{$p_2$}        & $\in (-5, 0)$ &  -1.9 & $10 \%$ 
\enddata
\end{deluxetable}

With our inferred surface density profile (power-law indexes $p_1 \sim 1.8, \, p_2 \sim -1.9$), the mass of the disk is strongly concentrated around $a_{\rm break} \sim 1 R_\odot$. 
We also find that $q\sim 2.4$, slightly steeper than our expectation for an optically thin disk ($q=2$) {\w and represents a disk that is radially optically thick to the ionizing photons.} It is worth commenting that, while one naively expects a degeneracy between $q$ and $(p_1, p_2)$, since all of them describe radial dependencies ($q$ on radius, while $p$ on semi-major axis), Fig. \ref{fig:corner} convincingly shows that the degeneracy is broken, likely due to the fact that the rings are substantially eccentric. 

\section{Rigidly Precessing Disk}
\label{subsec:precession}

Here, we first argue that the disk cannot be differentially precessing, as general relativity would have it. We then show, using the tool of linear eigen-mode calculations, that it has the appropriate gas pressure to resist  differential GR precession. In fact, both the observed precession period and the eccentricity profile agree with theoretical expectations.  
We thus firmly establish a long-suspected behaviour, that the J1228 disk is rigidly precessing.

We first establish a new estimate for the observed precession period. Previously, \citet{Manser2016} reported a period of 24-30 yrs, based on data up to May 2015. More recent monitoring extends the data to May 2018
\citep{2019Sci...364...66M} . From these, one infers that the triplet evolves through a symmetric profile around Oct. 2017. The last time they did so was around July 2007. This leads us to refine the precession period to a value of 20.5 yrs. 

We now also insert the factor of $1/\sin^2 i$ where necessary. Previously, \citet{2006Sci...314.1908G} have assigned an inclination of $i \sim 70^{\circ}$ to the disk, 
based on the crude arguments that the disk is far from being face-on (double-peaked emission), and is also not edge-on (no self-absorption). This value remains uncertain, so we keep it as a variable.

\subsection{GR Makes It Differential}

An eccentric ring that is in proximity of the white dwarf  experiences general relativistic (GR) precession. Let the complex eccentricity be 
\begin{equation}
E = e\,\exp(i {\varpi})\, ,
\label{eq:Edef}
\end{equation}
where $\varpi$ is the longitude of pericentre measured relative to a fixed direction in space.
GR acts to advance $\varpi$ at a rate 
\begin{eqnarray}
{\dot{\varpi}}_{\rm GR} & =& {\frac{3 G M_* \Omega}{c^2 a (1-e^2)}}  
\approx \frac{2\pi}{84\, {\rm yrs}}\, \frac{1}{(1-e^2)}\,
\left(\frac{a}{1 R_\odot}\right)^{-5/2}\,  ,
\label{eq:grrate}
\end{eqnarray}
where $\Omega = \left ( G M_*/ a^3 \right )^{1/2}$ is the Keplerian frequency. 

Since our gas disk extends radially  {\w (with a width $\Delta r \sim r$)}, 
the above equation suggests that the eccentric disk  should have been markedly twisted after only $10$ yrs, the precession period for the inner most orbit. If so, streamlines from different orbits could have crossed, and the resulting dissipation should have circularized and shrunk the disk. In contrast, the sharp spikes
 seen in the line profiles   suggest a significant eccentricity. In fact, the disk  has been observed to remain the same eccentric shape for over 20 yrs.

For this reason, previous studies have argued that J1228 and other similar white dwarfs do not harbour eccentric disks, but instead host  circular disks with non-axisymmetric brightness patterns \citep[e.g., spiral wave, vortex, ][]{Hartmann2011,Metzger2012}.There are also suggestions of eccentric disks but with misaligned apses \citep{Cauley2018,Fortin2020}. These proposals, while being able to provide reasonable fits to the data, are not  physically motivated. An asymmetric pattern on a circular disk can be rapidly sheared out on the Keplerian timescale (even faster than GR); and an eccentric disk with misaligned apses are not known to be self-sustaining. 

\begin{figure*}
    \centering
\includegraphics[width=0.9\linewidth]{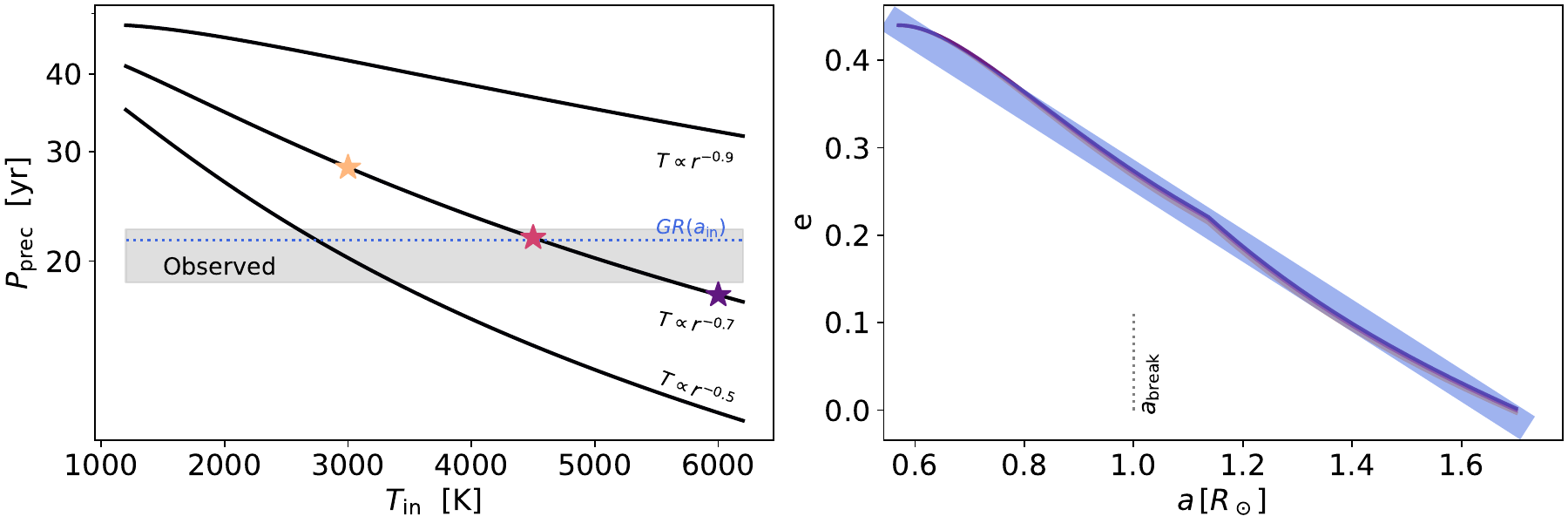}
\caption{Comparing our disk model to the linear eigen-mode calculations, assuming an inclination of $90\deg$. Solid curves in the left panel show the calculated precession period as a function of temperature at the inner edge, for a few choices of $\gamma$ (eq. \ref{eq:gamma}). The grey band corresponds to the precession period determined from data, and the blue dotted line that of GR period at the disk inner edge. The right panel shows the  calculated eccentricity eigen-functions in solid lines ({\w they are too close to tell apart)} corresponding to the three starred positions on the left panel. Our best-fit   eccentricity profile is plotted as the thick blue line. The theoretical modes are normalized to have the same eccentricity at the inner edge as the observed one. 
}
\label{fig:eigenmodes}
\end{figure*}

\subsection{Pressure Keeps It Rigid}

In our work, we opt to model the disk as a series of apse-aligned rings that rigidly precess. We now confirm that this is physically motivated.

First, the radial pressure gradient also causes precession, at a rate that is, to order of magnitude, 
\begin{equation}
    {\dot \varpi}_{p} \sim \left({{c_s}\over{v_{kep}}}\right)^2 \Omega \sim \frac{2\pi}{84 {\rm yrs}}
     \, \left(\frac{a}{1 R_\odot}\right)^{-3/2}\, \left(\frac{c_s/v_{\rm kep}}{0.0028}\right)^2 \, ,
\end{equation}
for a ring with  width $\Delta r \sim r$ \citep{Goodchild06,ogilvie2016}. To be competitive against GR (eq. \ref{eq:grrate}), we only need $c_s/v_{\rm kep} \geq 0.0024$, or a gas temperature\footnote{Cooler disks may still maintain rigid precession, but will require a very steep eccentricity gradient.}
\begin{equation}
    T \geq 1150 {\K} \times \left(\frac{\mu}{9}\right)\, ,
\end{equation}
where we have evaluated at $a = 1 R_\odot$ and have scaled the mean-molecular weight against that for singly-ionized metallic gas (see below).
The temperature of the gas disk is likely controlled by photo-ionization and ranges from $5000$ to $9000\K$ \citep{Melis2010}.{\w We have also confirmed this independently using the spectral synthesis code CLOUDY \citep{Ferland2017}.} So the whole disk can easily communicate via pressure, and can smooth out any precessional mis-demeanour. 

The negative eccentricity gradient we report here supports the hypothesis that the disk is rigidly precessing. Such a configuration means that the rings are more compressed together at their apocentres.
The radial pressure gradient there tends to precess the inner streamline backwards, while the outer streamlines precess forwards. This equilibrates their differential GR rates (eq. \ref{eq:grrate}). 

We now make the above arguments more quantitative.
We follow \citet{2018ApJ...857..135M} to compute the eccentricity eigenmode, the global coherent response of the disk to an eccentricity perturbation.  
\citet{ogilvie2016} have studied  the linear response of a locally isothermal, 3-D disk. For the case of a power-law disk, where the surface density scales as $\Sigma \propto r^p$, and where the temperature also obeys a power-law,
\begin{equation}
T (r) = T_{\rm in} \left( \frac{r}{r_{\rm in}}\right)^{-\gamma}\, , 
\label{eq:gamma}
\end{equation}
their result is simplified by \citet{2018ApJ...857..135M} into {\w a one-dimensional ODE,}
\begin{eqnarray}
& & \frac{\partial^2 E}{\partial r^2} + \frac{(3 - p)}{r} \frac{\partial E}{\partial r} + 
    \left [\frac{6 - \gamma(\gamma+2) - p(\gamma+1)}{r^2} \right. \nonumber \\  & & \left. + \frac{6r^2 \Omega^4}{c^2 c_s^2}- \frac{2 \Omega \omega_{prec}}{c_s^2}\right ] E = 0\, , 
\label{eqn:E_diff}
\end{eqnarray}
where $E = E(r)$ is the (apse-aligned) eccentricity eigenfunction, and
$w_{prec}$ the frequency of global precession. The isothermal sound speed is $c_s = \sqrt{k T/\mu m_H}$ and we adopt $\mu =9$ (see \S \ref{sec:insights}).

For our problem, 
since we only measure the length combination ${\tilde r} = r/ \sin^2 i$,
we transform the above equation to 
\begin{eqnarray}
& &\frac{\partial^2 E}{\partial {\tilde r}^2} 
+ \frac{(3 - p)}{\tilde r} \frac{\partial E}{\partial {\tilde{r}}} + 
    \left [\frac{6 - \gamma(\gamma+2) - p(\gamma+1)}{{\tilde r}^2}\right.
    \nonumber \\
    && \left. + \frac{6{\tilde r}^2 {\tilde \Omega}^4 }{c^2 c_s^2 \sin^4 i} - \frac{2 {\tilde \Omega} \omega_{prec}\sin i}{c_s^2} \right ] E = 0\, , 
\label{eq:realE}
\end{eqnarray}
where ${\tilde \Omega} = \sqrt(GM_*/{\tilde r})$ and $c_s^2 = c_{s0}^2 ({\tilde r}/{\tilde r}_{0})^{-\gamma}$.

One seems to have some flexibilities in choosing the boundary condition \citep[see, e.g.][]{2018ApJ...857..135M}.
We adopt the following set 
\begin{equation}
\left.\frac{\partial E}{\partial r}\right|_{\rm a_{\rm in}} =0 \, ; \hskip0.5in  E|_{\rm a_{\rm out}} = 0\, .
    \label{eqn:boundary}
\end{equation}
These differ from that in
\citet{2018ApJ...857..135M}, where they also took $\partial E/\partial r = 0$ at the outer boundary.  Our adopted one is more descriptive of our best-fit solution. In any case, this does not much affect the precession rate.

For our broken power-law disk, we integrate Eq. \ref{eq:realE} from the two boundaries towards ${\tilde a}_{\rm break}$ and insist that $E$ and $\partial E/ \partial r$ remain continuous across ${\tilde a}_{\rm break}$. We then look for eigenmodes of the lowest radial order. These have
the smoothest eccentricity profiles, and hence the lowest dissipation rates. {\w They are most likely to persist.} 

We present in Fig. \ref{fig:eigenmodes} the results of these calculations, adopting the best-fit model parameters from Table \ref{tab:bestfit}.As we do not have information on the values of $T_{\rm in}$ and $\gamma$, we experiment within some sensible ranges.
These calculations are performed for $i=90^o$, but the conclusion remains largely the same, given our uncertainties in $T_{\rm in}$ and $\gamma$, for inclinations as low as $\sim 60^o$.

We conclude {\w with} two findings. First, within the relevant range of $T_{\rm in}$, and sensible temperature profiles ($\gamma$), we find that the theoretical modes have periods comparable to the observed value ($20.5$ yrs). Second, even though we have only modelled the disk using a linear profile (the simplest choice), this profile agrees very well with the shape of the linear eigen-functions.  
 These two quantitative agreements strongly supports the hypothesis that J1228 hosts a rigidly precessing gas disk, under the combined effects of GR and gas pressure.

\section{Insights on Origin}
\label{sec:insights}

We now discuss what our results imply for the origin of the gas disk around J1228, as well as for the pollution of white dwarfs in general. We base our discussions on the current favourite model for white dwarf pollution, tidal disruption of highly eccentric planetesimals {\w \citep[see reviews by][]{2016NewAR..71....9F,2022MNRAS.509.2404B}}.

\subsection{Disk Mass and the Progenitor Mass}

We estimate a mass for the gas disk, $M_{\rm gas}$, by assuming that viscous spreading of the gas disk supplies the observed accretion onto J1228.

The gas scale height is 
\begin{equation}
    \frac{H}{r} = \frac{c_s}{v_{\rm kep}} \sim 0.006 \left({\mu\over{9}}\right)^{-1/2}\left(\frac{T}{5000 {\rm K}}\right)^{1/2}\, \left(\frac{r}{1 R_\odot}\right)^{1/2}\, .
    \label{eq:hoverr}
\end{equation}
If the disk is accreting under a constant viscosity parameter $\alpha$ \citep{SSalpha}, the viscous diffusion time is 
\begin{eqnarray}
{\tau_{\rm diff}} & \approx & { \alpha^{-1} \Omega^{-1} \left(\frac{H}{r}\right)^{-2}}
\nonumber \\
&\sim &
200 {\rm yrs}\, 
\left(\frac{r}{1 R_\odot}\right)^{3/2}\, 
\left(\frac{\alpha}{10^{-2}}\right)^{-1}\, 
\left(\frac{H/r}{0.006}\right)^{-2}\, 
\, . 
\label{eq:tdiff}
\end{eqnarray}
We adopt the accretion rate as determined by \citet{bauer2023}\footnote{This updated rate accounts for diffusion from thermohaline mixing and is much greater than the estimate of in $\dot M = 5.6\times 10^{8}\g/\s$ in \citet{gaensicke2012}, .} and estimate the current disk mass by assuming ${\dot M} \sim M_{\rm gas}/\tau_{\rm diff}$, to obtain
\begin{eqnarray}
M_{\rm gas} & \approx & 2\times 10^{21} {\rm g} \left(\frac{\dot M}{2\times 10^{11} {\rm g}/{\rm s}}\right) \left(\frac{\alpha}{10^{-2}}\right)^{-1} \nonumber \\
& & \times \left(\frac{T}{5000{\rm K}}\right)^{-1} \left(\frac{r}{1 R_\odot}\right)^{1/2}\, .
\label{eq:mgas1}
\end{eqnarray}

We argue that this estimate likely also reflects the total mass of the original disrupted planetesimal. The current disk is significantly  eccentric --  it avoids rapid streamline crossing 
and circularization by organizing itself into a coherent eccentric mode.  But over the viscous timescale,  the eccentricity should be gradually damped as the disk spreads radially. So the current disk has likely weathered no more than a few viscous times. Or, its current mass is close to its original mass. If true, this means the original disrupted plantesimals has a radius {\w $R_p \sim 50\km$ (assuming a bulk density of $\rho_p \sim 5\g/\cm^3$).}


To substantiate the above estimate for the disk mass, we also consider whether it is consistent with the observed Ca II emissions. Let us adopt the same chemical composition as that for CI chrondrite from \citet{palme2014}, 
and assume that all metals are singly ionized, the mean nuclei weight of the gas is $15$ and the mean molecular weight is $9$ {\w (in unit of hydrogen mass)}. So the above mass estimate corresponds to a surface density $\Sigma \sim M_{\rm gas}/r^2 \sim 
0.13 \g/\cm^2$, and a midplane density of $\rho \sim 3\times 10^{-10} \g/\cm^3$.
In the meantime,  Ca nuclei are only $1/280$ of the total nucleus number.
We therefore arrive at a radial column density for Ca of
$\sim 3\times 10^{21} \cm^{-2}$, and a vertical one at $\sim 2\times 10^{19} \cm^{-2}$. We find:

\begin{itemize}
\item if all Ca is in Ca II,\footnote{Ca I, with an ionization potential of $6.11$eV, is easily ionized; while Ca II, with a potential of $11.87$eV, is much harder to ionize.},  with a photo-ionization cross-section of $5\times 10^{-19} \cm^{2}$,
the optical depth for a Ca II ionizing photon 
is $\tau \sim 1600$. 
So the gas midplane is very opaque to  the Ca II ionizing photons. This may explain why we obtain a steeper fall-off of the ionizing flux ($n_\gamma \propto r^{-2.4}$), than is expected for an optically thin disk ($n_\gamma \propto r^{-2}$).

\item the vertical optical depth for the Ca II infrared triplet is $\sim 10^6$ ({\w CLOUDY} result). This explains why the line ratios within the CaII triplet do not reflect their individual strengths, but approach those of a blackbody \citep{Melis2010}.

\item the above mass estimate allows us to explain the total flux observed in Ca II triplet. For J1228, about $5.6\% $ of its energy are in photons that can ionize Ca II (ionization energy $11.87$ eV). Let the disk be optically thick to these photons from the midplane up to $n$ scale heights. The total energy intercepted by Ca II is then $2 n \times (H/r) \times 5.6\%\times L_{\rm wd}$. As Ca II is ionized and recombined, a fraction of the ionization energy is emitted in Ca II triplets (photon energy $\sim 1.5$ eV). So we expect a total line flux of $\sim 2 n (H/r) \times 5.6\% \times (1.5/11.87) \times L_{\rm wd}$. The observed total line flux is $\sim 3\times 10^{-4} L_{\rm wd}$ \citep{Manser2016}, or we require $n \sim 3.5$. In comparison, using the above midplane radial optical depth ($\tau \sim 1600$), and assuming that the disk is in vertical hydro-static equilibrium, we find that the disk can capture these photons up to a height multiple of $n\sim 3.8$.  In other words, the observed CaII line fluxes can be explained by our inferred disk density.
\end{itemize}

\subsection{The Three Radii}
\label{subsec:insights1}

We aim to draw clues on the progenitor by considering the three radii we inferred for the disk, $a_{\rm in}$, $a_{\rm break}$ and $a_{\rm out}$.

With our values of $p_1$ and $p_2$, most of the disk mass lies closely around $a = a_{\rm break} \simeq 1\times \sin^2 i
\, R_\odot$. This suggests that $a_{\rm break}$ may be a special radius for the progenitor body. Gas deposited here may then viscously spread both outwards and inwards, forming the extended disk. In particular, \citet{Metzger2012,2016ApJ...830....7R} showed that the surface density of an isothermal accretion disk should scale as $r^{-2}$, similar to our value of $p_2 = -1.9$.

We then consider two physical radii, one for dust sublimation and the other for tidal disruption. 

Using the most updated stellar parameters from 
\citet{koester2014},
 $M_*= 0.705 M_\odot$, $R_* = 0.0117 R_\odot$, $T_{\rm eff} = 20,713{\rm K}$,
 a blackbody at  distance $r$ is heated to a temperature 
 \begin{equation}
 T_{\rm bb} = 
 \left(\frac{1}{4}\right)^{1/4} \left(\frac{R_{\rm wd}}{r}\right)^{1/2} T_{\rm wd}\sim 
1300 {\rm K}\,  \left(\frac{r}{1.5 R_\odot}\right)^{-1/2}\, ,
     \label{eq:tbb}
 \end{equation}
 where $1300 {\rm K}$ is roughly the sublimation temperature of silicate grains under our disk midplane density, $\rho
\sim 3\times 10^{-10}\g/\cm^3$ \citep{Pollack94}.\footnote{The fit provided by \citet{Isella} is    $T_{\rm evap} 
\sim 1600 \, {\rm K} \left(\frac{\rho}{10^{-5} {\rm g}{\rm cm}^{-3}}\right)^{0.0195}$.}
This will place the sublimation radius at around $1.5 R_\odot$,  much beyond {\w our inferred break radius} ($a_{\rm break} = 1 \sin^2 i \, R_\odot$), but close to our inferred outer edge {\w for the gas disk}, $a_{\rm out} = 1.7 \sin^2 i\,  R_\odot$. It is therefore likely that the gas disk is truncated near the sublimation boundary. There may well be a dust component lying beyond it (see discussions below).

On the other hand, the tidal disruption radius {\w typically lies closer than the sublimation radius}. For a body with negligible internal strength\footnote{A finite internal strength will allow the body to survive closer to the white dwarf \citep[see, e.g.][]{Zhang2021}.}
and on a parabolic orbit ($e\approx 1$), {\w this} is located at a peri-centre distance of \citep[][]{sridhartremaine,watanabe1992}
\begin{eqnarray}
\left.r_{\rm roche}\right|_{e\approx 1} &\approx & 1.7 R_{\rm wd} \times \left({{\rho_{\rm wd}}\over{\rho_{\rm p}}}
\right)^{1/3} \, \nonumber \\
& & \approx 1.0 R_\odot \left(\frac{\rho_{\rm p}}{5\, \g/\cm^3}\right)^{-1/3}
\, ,
\label{eq:roche}
\end{eqnarray}
{\w where $\rho_p$ is the body's bulk density.}
{\w If the orbit is less eccentric, this radius moves outwards,  reaching a maximum of $1.5 R_\odot$ at $e=0$ \citep{Chandra}.}

{\w Given our observed break radius ($a_{\rm break} = 1 \sin^2 i R_\odot$), it seems reasonable to suggest that 
the progenitor enters the zone of tidal disruption, is threaded apart, and its debris then sublimates to form the observed disk \citep[also see][]{McDonald-Veras21}.
There is much energy dissipation during this process, so the resultant disk is expected to be more compact and less eccentric. This so-called 'circularization' process has been discussed by 
\citet{2015MNRAS.451.3453V,2020MNRAS.498.4005O, 2021MNRAS.501.3806M} and
\citet{2021MNRAS.505L..21T}. Of particular note is the \citet{2021MNRAS.505L..21T} work, where they  investigated the formation of a gaseous disk  under an eccentric, out-gassing parent body. Unfortunately, current simulations can only integrate for hundreds of orbits, or in physical time, $\sim$ days. This is too short to predict the final disk shape (eccentricity and mass) which evolves on precession timescales (decades).}

Lastly, the observed line profiles clearly indicate that the gas disk has an abrupt inner cutoff at $a_{\rm in} \sim 0.57 \sin^2 i \, R_\odot $. {\w What is the origin of this cutoff?}
\citet{Metzger2012} suggested that stellar magnetic field may be able to truncate the disk, much like that around in T Tauri stars. However, we suspect a different explanation may be at work here.   
In the inner part of the disk, rigid precession under the stronger GR precession demands a steeper eccentricity gradient. If the disk extends closer to the white dwarf than is observed, the  implied high eccentricity will be challenging for its survival --
nonlinear effects may  disrupt the pattern of rigid precession and cause streamline crossing.   
We hypothesize that disk eccentricity, rather than stellar magnetic field, truncates our disk at the observed $a_{\rm in}$. More detailed study is required.

\subsection{A Co-spatial Dust Disk?}
\label{subsec:insights4}

The spectral energy distribution of J1228 shows the presence of a dusty component, with a total luminosity of $L_{\rm dust}\sim 6\times 10^{-3} L_{\rm wd}$, and blackbody temperatures that range from $450\K$ to $1700\K$ 
\citep{2009ApJ...696.1402B}. If the dust lies in a geometrically flat disk and sees the central star unobstructed, it should be illuminated to a temperature \citep{ChiangGoldreich,Jura2003}

\begin{eqnarray}
T_{\rm dust} &=& \left({2\over{3\pi}}\right)^{1/4} \left(\frac{R_{\rm wd}}{r}\right)^{3/4} T_{\rm wd} \nonumber \\
& & \sim 350 \K \left(\frac{r}{1.5 R_\odot}\right)^{-3/4}\, .
\label{eq:tdust}
\end{eqnarray}

So the above dust temperatures translate to a range of $0.2 R_\odot$ to $1.2 R_\odot$.  

But {\w such a model} would place the dust component in the same 
{\w annulus} 
as the gas disk. The eccentricity of the gas disk makes this problematic. The gas and dust components, if orbiting at different eccentricities and precessing independently (i.e., dust experiences GR but not gas pressure), will encounter each other at enormous speeds,  of order a few hundred km/s. This {\w would} lead to {\w circularization of the gas disk, if the dust mass is high enough.}
In fact, this un-welcomed prospect led \citet{Metzger2012} to suggest that the gas disk cannot be eccentric, a proposition now amply refuted by our analysis.

One way to resolve this is if the {\w dust component do not lie in a flat, opaque disk, a proposal that is also supported by dust observations of GD 56 \citep{2007ApJ...663.1285J}, WD J0846+5703, \citep{2021MNRAS.504.2707G} and G29-38 \citep{2022ApJ...939..108B}.} If the dust 
grains are not in a flat disk and do not block each other's view to the star,
their temperaturets will then be described by eq. \ref{eq:tbb}, {\w and they can remain hot out to larger distances.} The observed blackbody may then arise from a region beyond  $0.9 R_\odot$, largely avoiding the most eccentric part of the gas disk. These grains can lie even further away, if they are smaller than the wavelengths of their own thermal radiation and are thus super-heated.

Such a situation (free-floating grains) can arise if the grains are short-lived and have not yet undergone collisional flattening (into a thin disk). Because of the proximity of the gas disk to the sublimation radius, these grains may be in condensation/sublimation equilibrium with the gas disk, and are transiently formed and destroyed. 
In this case, one can obtain a lower-limit to the dust mass by assuming that the observed dust luminosity is produced by grains of size $s$, bulk density $\rho_{\rm bulk}$ and temperature $T$, 
\begin{eqnarray}
   M_{\rm dust} & \geq & \frac{L_{\rm dust} \,  s \, \rho_{\rm bulk} }{3 \sigma T^4} \sim  2\times 10^{17}\g \, \, \left(\frac{L_{\rm dust}}{6\times 10^{-3} L_{\rm wd}}\right) \times \nonumber \\  & & \left({T\over{1600\K}}\right)^{-4} \left(\frac{s}{1\mu m}\right)
 \left(\frac{\rho_{\rm bulk}}{5\g/\cm^3}\right)\, . \label{eq:Mdust}
\end{eqnarray}
In other words, only a minute amount of dust is needed to reproduce the observed SED.
{\w Most of the progenitor mass may instead lie in the gas disk.} 

In this scenario, the dust component (and its Poynting-Robertson drag) will be irrelevant for the evolution and accretion of the gas disk, differing from the proposal by \citet{2011MNRAS.416L..55R,Metzger2012}. 

\subsection{The Progenitor}
\label{subsec:insights3}

Here, we remark on how the J1228 disk informs on the origin of white dwarf pollution.

The observed gas disk is markedly  eccentric. Barring the possibility that the eccentricity is excited after formation \citep[see, e.g., proposal from][]{2018ApJ...857..135M}, this points to a very  eccentric orbit for the progenitor. This  is expected in the hypothesis of tidal disruption. 

We can infer the original peri-centre approach, by assuming that the orbital angular momentum ($\sqrt{a(1-e^2)}$) is largely conserved when the tidal debris is circularized into the observed disk. This yields a pericenter distance, $r_p = a_{\rm orig} (1-e_{\rm orig}) \sim a_{\rm break}*(1-e_{\rm break}^2)/(1+e_{\rm orig}) \sim 0.47 \sin^2 i\,  R_\odot$,  for $e_{\rm orig} \approx 1$ and $e_{\rm break} \sim 0.26$ at $a_{\rm break}$. This is a couple times smaller than the Roche radius (eq. \ref{eq:roche}). 

Who can place the progenitor on such an odd orbit, with an implausibly small pericenter approach? The most likely scenario left is planetary perturbations {\w \cite[see review by][]{Veras2021}. One often invoked mechanism is the so-called Kozai-Lidov oscillations \citep{Kozai1962,Lidov1962}, wherein secular torque from a strongly misaligned planetary perturber kneads the progenitor orbit into an almost needle-like radial orbit.\footnote{Compared to the Kozai-Lidov process, other known dynamical channels do not favour such extreme eccentricities and have much smaller probabilities of success.} However, such a process could be easily suppressed by GR precession \citep[e.g.][]{Eggleton2001}. To avoid the suppression,}  the secular perturbations may need to act in cohort with  mean-motion interactions and/or close encounters {\w \citep[the so-called 'planet scattering', see, e.g.,][]{Nagasawa2008}}.
This then requires that the orbital separation of the progenitor to be within a factor of few from that of the planet. 

{\w Where could the putative planet lie? We invoke the following estimate to argue that it must not be much further than an AU. Such a new constraint, un-seen before in literature, is enabled by the very short diffusion time of the gas disk. Things have to happen in a hurry.

After the progenitor is shredded apart, its debris fly on different orbits, but ones that have long axis well aligned with the original one. They then start on the process of circularization. In principle, differential GR precession can scramble the orbital orientations and produce crossing orbits, which then circularize as mutual collisions dissipate the kinetic energy. 
However, if the progenitor's semi-major axis is $a$, and its pericentre distance $r_p \sim 0.47 R_\odot$, the GR precession time is of order $\sim 2.5\times 10^5 (a/{\rm AU})^{3/2}$ yrs (eq. \ref{eq:grrate}). 
This is so long, it seems hard to circularize the  debris disk before the gaseous component diffuses away (eq. \ref{eq:tdiff}, $\sim 200$ yrs). In other words, GR alone will not circularize the debris disk fast enough to allow much gas accumulation.

On the other hand, if a planetary perturber lies within an AU, it is able to quickly scramble the debris orbits. We come to this conclusion as follows. To reduce the orbital energy by order unity (to achieve an order unity shrinkage of the semi-major axis), the debris orbits need to be misaligned by of order unity. This can be induced either by secular perturbations from the planet, or by successive scatterings. Both take about the same time for a particle that crosses the planet's orbit. So we will simplify the argument by setting the scrambling time to be the secular time, 
\begin{equation}
T_{\rm secular} \approx \left({{M_*}\over{M_{\rm planet}}}\right) P_{\rm planet}\, .
\end{equation}
This is consistent with results from \citet{Li2021}, where a Neptune-massed planet at 10AU takes $\sim 10^6$ yrs to scramble the orbits.

To replenish the gas disk within its short diffusion time ($200$yrs), we therefore require the scrambling time to be shorter than the diffusion time, or
\begin{equation}
a_{\rm planet} \leq 0.3 {\rm au}\, \left({{M_{\rm planet}}\over{M_{\rm Jup}}}\right)^{2/3}\, ,
\end{equation}
where $M_{\rm Jup}$ is the mass of Jupiter.
}

{\w Such a close-in planet is at tension with planet survival around an evolved star. Previous calculations \citep{2009ApJ...705L..81V,2011ApJ...737...66K,2012ApJ...761..121M,2013ApJ...777L..30A,2013MNRAS.432..500N,
2014ApJ...794....3V,
2016MNRAS.463.1040M,2020ApJ...898L..23R} have shown that as a star evolves through the giant phases and then loses its envelope, the closest distance one expects a planet (or planetesimals) to survive should lie outside a few AU -- reaching $\sim 7$au for an initial stellar mass of $3M_\odot$ \citep[Fig. 10 of][]{2012ApJ...761..121M}.

It is unclear yet how to resolve this tension. If the perturber lies much further away, one has to come up with an alternative way to rapidly circularize the debris disk. Past proposals have included radiative drag \citep{Veras+15} or collisional drag with a pre-existing disk \citep{Malamud+21}. However, these are not effective unless the debris have been ground down to microscopic sizes, and even then it is hard to satisfy the $200$ yr constraint. }

{\w We now turn to  the source population that gives rise to the progenitor.

If the gas disk disappears in a few diffusion time ($\sim 200 $yrs), that means the disruption event must have occurred only recently, within the past few centuries. 
We then estimate how common such an event occurs during the lifetime of J1228.
White dwarfs that are of  a similar temperature as J1228 (so that they can sublimate grains and maintain a gaseous accretion disk) have a space density of $\sim 4\times 10^{-5}/{\rm pc}^3/{ \rm mag}$ \citep{leggett1998}. 
So within the distance of J1228 ($\sim 127$pc), 
there are about $300$ such white dwarfs. Among these, about two dozen gas disks have been reported \citep{2006Sci...314.1908G, 2007MNRAS.380L..35G, Gansicke2008,2012MNRAS.421.1635F,Fusillo2021}. So the occurrence rate of gas disks among hot white dwarfs is  $\sim 10\%$, not dissimilar to the occurrence rate of dust disks among all white dwarfs.\footnote{We note that \citet{Manser2020} estimated an occurrence rate of $0.067\pm ^{0.042}_{0.025}\%$ for gaseous disks among alll white dwarfs. This is comparable to our estimate here, since hot white dwarfs like J1228 are of order $1\%$ of the local white dwarf population\citep{leggett1998}.} So the total mass of (disrupted) progenitors (all assumed to be $R_p=50\km$) over the cooling age of J1228 ($\sim 10^7$ yrs) \footnote{Here, we ignore events that may have occurred during the main sequence phase.} is,
\begin{eqnarray}
M_{\rm disrupted} & & \approx {{4\pi} \over 3} \rho_{p} R_p^3 \, \times {{{10\%} \times {10^7 {\rm yrs}}}\over{t_{\rm diff}}} \nonumber \\
& & \sim 10^{25}\g\, \times \left({{R_p}\over{50\km}}\right)^3 \left({{\rho_p}\over{5\g/\cm^3}}\right) .
\label{eq:Mtotal}
\end{eqnarray}
This mass is similar to the total mass of our own asteroid belt ($\sim 2\times 10^{24}\g$). However, the disrupted bodies are those that are able to reach the Roche radius, and so likely only constitute a small fraction of the total source disk. Moreover, evidence on older white dwarfs suggests that metal pollution continues well past the cooling age of J1228. Both these arguments imply that, for us to witness a rapidly draining gas disk around J1228, there must exist planetesimal disks around these white dwarfs that are much more massive than what one expects from their main-sequence counter parts. This, {a massive belt at au distance,} is again at tension with our current knowledge.} 
{While a planetesimal belt at the Kuiper belt distance can remain massive for longer, due to a slower collision timescale, a belt at the Asteroid belt distance should grind down quickly during the main-sequence lifetime of a star, even if it started massive \citep[see, e.g.][]{Wyatt2007}. }



{\w Lastly, we  comment on the report of a planetesimal embedded in the gas disk.  \citet{2019Sci...364...66M} discovered  periodic variations in the line profiles of J1228, with a period of $123.4$min (corresponding to an orbital semi-major axis of $0.73 R_\odot$). They interpreted these disturbances as due to an out-gassing planetesimal on an eccentric orbit that overlaps with the gas disk. If such an object does exist {\bf and} is massive (radius $R \geq 50\km$), our above arguments on the disk origin can fail, because the disk can now be continuously fed and can survive much longer than the diffusion time.\footnote{Our main results on disk profiles and precession properties remain valid.} However, 
there are multiple reasons to suspect the planetesimal interpretation. First, as remarked in \citet{2019Sci...364...66M}, the survival of such a body well inside the nominal Roche radius is problematic:  the body has to have an unphysically high bulk density ($\rho_p \geq 40$g/cm$^3$ for circular orbit; higher value needed for eccentric orbit; see eq. \ref{eq:roche}); or the body has to have a high internal strength $\sim 10^8  (R_p/50\km)^2
\, \dyne/\cm^2$ \citep[see eq. 4.78 of][]{MurrayDermott}, comparable to that of a monolithic rock.
Moreover, if the planetesimal precesses at a different rate than the gas disk (which is physically likely), there will be continuous streamline crossings between the existing gas disk and the newly sublimated gas, resulting in a disk that is nearly circular and much more compact than the parent body. Lastly, it is hard to imagine how a massive planetesimal can evolve from a nearly parabolic orbit to its current close orbit, short of any sensible dissipation mechanism.

}

\section{Conclusion} \label{sec:conclusion}

We undertake a detailed modelling of the gaseous disk around the white dwarf SDSS J1228+1040. We find that the disk has a surface density profile that peaks around $1 \, \sin^2 i \, R_\odot$, and an eccentricity profile that decreases outward. The latter, we show, uncannily reproduces the theoretical profile of a disk that precesses rigidly 
under the combined forces of general-relativity and gas pressure. In other words, the observed disk is in an eccentric eigen-state. This explains why the disk can be eccentric yet long-lived. As we expect the eccentricity to be dissipated in the viscous timescale ($\sim 200$ yrs), the current disk should have formed fairly recently. 

Based on the high accretion rate onto the white dwarf (current estimate $\dot M \sim 2\times 10^{11} \g/\s$), 
we infer a mass of $\sim 10^{21} \g$ for the gaseous disk. Such a mass estimate is also consistent with the emission measures in the Ca II triplets. The young age of the current disk then implies that it still contains most of the source mass, pegging the progenitor at a size of $R \sim 50\km$.

Given the eccentricity of the gas disk, there is unlikely to be a massive dusty debris disk that is co-spatial. Rather, we suggest that the observed dust emission may arise from small amounts of grains that are in condensation/sublimation equilibrium with the gas disk. 

{\w The eccentricity of the remnant disk supports the hypothesis of tidal disruption. The progenitor body, after being excited to a nearly parabolic orbit and reaching a peri-centre distance of $0.47 \sin^2 i R_\odot$, was tidally disrupted by the white dwarf. The debris then circularizes into the current moderately eccentric and compact disk. 

The short diffusion time of the gas disk allows us to set unique and stringent constraints on the origin of the system.  By insisting that the tidal debris circularizes into the current size in a time shorter than the gas disk diffusion time, we find that both the planetesimal and its planetary perturber likely orbit around the white dwarf within an AU. This is in tension with our current knowledge of planet survival around white dwarf progenitors.}

Moreover, to account for the observed rate of gaseous disks among similarly hot white dwarfs, we estimate that the source disk needs to contain 
{\w at least the mass of our asteroid belt, and likely many orders of magnitude more in reality. Such a massive disk does not seem to exist around the main-sequence counter-parts of these white dwarfs.}

Our study in this work is preliminary in nature. We have not investigated in detail the temperature structure and the emission mechanism of the metallic disk. Emission line diagnostics may be used to constrain the disk inclination, which may lead to further insights.  We also fall short of analyzing the `circularization' process after tidal disruption. This latter seems a promising route to infer the nature of the progenitor and the architecture of planetary system around J1228.

Such careful studies are clearly warranted. J1228 is likely not unusual. First, 
many white dwarfs with gaseous disks show variable emissions, indicating eccentric, precessing disks \citep{Gansicke2008,Melis2010,2014MNRAS.445.1878W,Cauley2018,Manser2021}. Similar dynamics as we reveal here for J1228 may be in play in all of these disks. Second, while J1228 is hot and can sublimate rocks at around $1 R_\odot$, cooler white dwarfs will only harbour fully dusty disks. Such a disk reveals no information on its kinematics and surface density. J1228 offers us a lucky window into these otherwise obscure disks and may well be the Rosetta stone to  decipher the mystery of white dwarf pollution.

\bigskip
We acknowledge NSERC for funding. We also thank C. Manser for providing the line data,  Renu Malhotra and Elliot Lynch for discussions. Lastly, we thank the anonymous  referee for a careful critique of the paper.

\bibliography{gasdisk}
\bibliographystyle{aasjournal}

\end{document}